\documentclass[conference,10pt]{IEEEtran}
\IEEEoverridecommandlockouts

\usepackage{cite}
\usepackage{caption}
\usepackage{amsmath,amssymb,amsfonts}
\usepackage{algorithmic}
\usepackage{graphicx}
\usepackage{textcomp}
\usepackage{xcolor}
\usepackage{url}
\usepackage{colortbl}
\def\BibTeX{{\rm B\kern-.05em{\sc i\kern-.025em b}\kern-.08em
    T\kern-.1667em\lower.7ex\hbox{E}\kern-.125emX}}

\usepackage{hyperref}
\hypersetup{
    colorlinks=true,
    urlcolor=magenta,
}

\begin{document}

\title{Soft Disentanglement in Frequency Bands for \\ Neural Audio Codecs\\
\thanks{This work was funded by the European Union (ERC, HI-Audio,
101052978). Views and opinions expressed are however those of the author(s) only and do not necessarily reflect those of the European Union or the
European Research Council. Neither the European Union nor the granting
authority can be held responsible for them.}
}

\author{
    \IEEEauthorblockN{
    Benoît Giniès,
    Xiaoyu Bie, 
    Olivier Fercoq, 
    Gaël Richard
}
\IEEEauthorblockA{
    LTCI, T\'el\'ecom Paris, Institut polytechnique de Paris, Palaiseau, France\\
    }
}

\maketitle

\begin{abstract}
In neural-based audio feature extraction, ensuring that representations capture disentangled information is crucial for model interpretability. However, existing disentanglement methods often rely on assumptions that are highly dependent on data characteristics or specific tasks. In this work, we introduce a generalizable approach for learning disentangled features within a neural architecture. Our method applies spectral decomposition to time-domain signals, followed by a multi-branch audio codec that operates on the decomposed components. Empirical evaluations demonstrate that our approach achieves better reconstruction and perceptual performance compared to a state-of-the-art baseline while also offering potential advantages for inpainting tasks.
\end{abstract}


\section{Introduction}

In audio signal processing applications, including feature extraction, recognition, and synthesis, an efficient representation of the raw signal is essential. The primary goal of feature representation is to encode the most relevant information in a compact form. Traditionally, this trade-off between data completeness and representation efficiency has led to the development of handcrafted features designed to capture meaningful properties of the signal. In this context, time-frequency representations play a crucial role, with widely adopted approaches ranging from the Short-Time Fourier Transform (STFT) to perceptually motivated representations such as the mel scale and the constant-Q transform \cite{stevens1937scale, Brown1991CalculationOA, wang_hybrid_2019}.

Efficient signal representation is particularly critical in audio compression, where the objective is to achieve high-fidelity reconstruction under strict bitrate constraints. Significant advancements have been made in this field since the mid-1990s, particularly with the development of standardized MPEG audio codecs \cite{mpeg2aac,mpeg4he-aac,mpegh} (see \cite{audio21stcentury} for an overview). More recently, the integration of neural networks has fundamentally reshaped this task, leading to state-of-the-art models that achieve remarkable compression rates with high audio quality \cite{zeghidour_soundstream_2021, defossez_high_2022, kumar_high-fidelity_2023}. Most of these approaches adopt an encoder-decoder architecture, which inherently serves as a bottleneck for feature extraction and is further coupled with a quantization process \cite{kingma_auto-encoding_2022}. 

While these algorithms achieve high compression efficiency, the extracted features convey little to no semantic information. This limitation arises because neural codecs, including VQ-VAEs, are typically trained without explicit constraints that promote interpretability in the learned representations. To address this issue, various research efforts have focused on enhancing the semantic relevance of neural features by structuring models to capture specific acoustic properties. For example, prior work has aimed to disentangle speech characteristics \cite{polyak2021speech, ju_naturalspeech_2024} or separate audio from noisy observation~\cite{omran_disentangling_2023} or audio mixture~\cite{bie2024learning}. This process, known as feature disentanglement, has been explored in a wide range of applications \cite{takahashi_hierarchical_2021,luo_gull_2024}.

In this work, we introduce a novel neural codec\footnote{Examples and code available at: \url{https://soft-disentangled-codec.github.io/}} that generates a discrete, frequency-disentangled representation. The proposed codec model operates across multiple sampling frequencies, enabling the extraction of a representation composed of discrete tokens, each corresponding to predefined frequency bands. We demonstrate that this disentangled representation yields a modest improvement over the baseline model and opens new possibilities for sound transformation tasks. As an illustration of its potential, we present a simple inpainting experiment.

\begin{figure*}[htbp]
\centerline{\includegraphics[width=1.5\columnwidth]{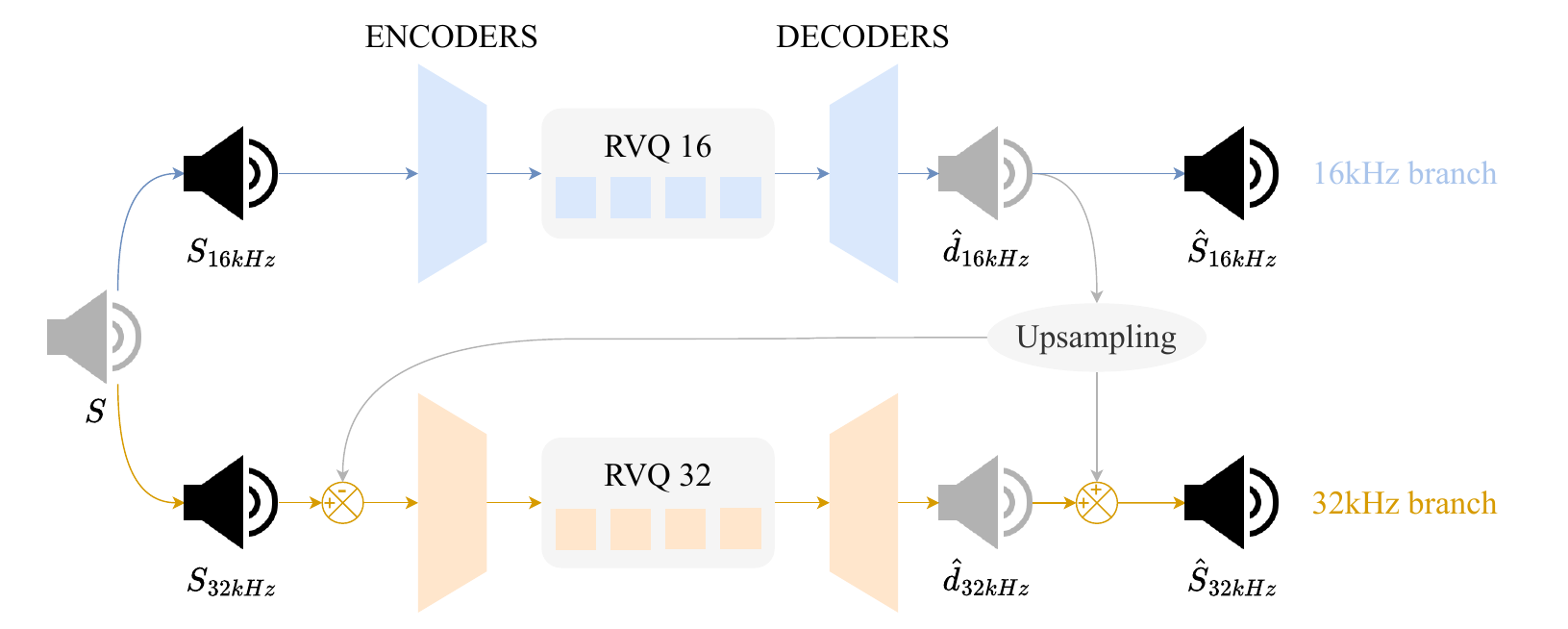}}
\captionsetup{justification=centering}
\caption{Proposed disentangled codec.  
The $16~kHz$ branch reconstructs the $[0-8~kHz]$ signal. The $32~kHz$ branch processes the residual of $S_{32kHz}$ and $U(\hat{d}_{16kHz})$ to output the $[0-16~kHz]$ signal summing the outputs from each branch.}

\label{fig-Um}
\end{figure*}

\section{Related Work}
\label{sec-related}

\subsection{Neural Audio Codecs}

The use of neural networks for feature extraction in image and audio processing gained significant traction with the introduction of the encoder-decoder architecture, initially applied in the VAE model \cite{kingma_auto-encoding_2022} and later extended with an additional quantization step in the Vector Quantized VAE (VQ-VAE) model \cite{oord_neural_2018}. This quantization step was further refined to enhance compression, leading to the Residual Vector Quantizer (RVQ) used in \cite{zeghidour_soundstream_2021, defossez_high_2022, kumar_high-fidelity_2023}. A relaxation of quantization techniques was explored in \cite{takida_sq-vae_2022}, and various quantization structures have been developed \cite{takida_hq-vae_2023, williams_hierarchical_2020}. Notably, the ability to access discrete representations has broader applications beyond audio compression, including speech synthesis \cite{wang_neural_2023}, music timbre transfer~\cite{Cifka2021}, and sound manipulation \cite{Bazin2021}.

\subsection{Feature Disentanglement}

The success of neural audio codecs has spurred growing interest in feature disentanglement, with methods often tailored to specific applications. For instance, in singing voice conversion, Takahashi \& al. \cite{takahashi_hierarchical_2021} integrate the VQ module with pitch and amplitude encoders. In speech synthesis, Adam \& al. \cite{polyak2021speech} enforce the disentanglement of content, pitch and speaker identity through specialized feature extraction structures, while Ju \& al. \cite{ju_naturalspeech_2024} allocate VQ modules to capture speech prosody, context, and acoustic details, complemented by an additional timbre encoder. The training of these vector quantizers is guided by appropriate supervision tasks. This idea of constraining disentanglement via supervised tasks has also been applied to speech-noise separation \cite{omran_disentangling_2023} and source separation~\cite{bie2024learning}. While many of these approaches are application-specific, some more general strategies have been proposed. For example, \cite{hsu_disentanglement_nodate} outlines a method to encourage self-disentanglement of quantized features, and Luo \& al. \cite{luo_gull_2024} introduce a generic approach with a multi-band codec, where each frequency band of the input spectrogram is processed separately.

\section{Disentangled Codec}
\label{sec-DC}

In this paper, we present a generic approach applicable to various types of audio signals. Our method leverages the inherent interpretability of frequency decomposition in audio signals, with the goal of obtaining a discrete feature representation that preserves frequency information. The core of our approach involves adapting the architecture of a discrete neural audio codec to reconstruct spectral information from the extracted discrete representation. One straightforward strategy to enforce frequency-dependent analysis is to split input spectrograms into sub-bands, which are then processed independently (as in Luo \& al. \cite{luo_gull_2024}). However, this method imposes significant constraints, as each band must be reconstructed separately in parallel with the others. Additionally, combining the sub-bands back into a full spectrogram may introduce undesirable artifacts.

Our approach does not suffer from these drawbacks since it relies on a soft frequency decomposition. We indeed
rather work in the temporal domain, by exploiting the link between a signal sampling rate and its spectral information. 
In this paper, we only consider a simple version of our approach with only two separated branches (see Fig. \ref{fig-Um} for a schematic description). 
Although simple, this two branches architecture can serve as a proof of concept that can readily be extended to multiple branches each operating at different sampling frequencies. The two branches of our neural audio codec operate at different sampling rates, respectively at $16~kHz$ and $32~kHz$. The input time domain signals $(S_{16kHz},S_{32kHz})$ are two versions of the same signal $S$, where $S_{16kHz}$ is sampled at $16~kHz$ (i.e. containing spectral information in the $[0-8~kHz]$ band) and $S_{32kHz}$ is sampled at $32~kHz$ (i.e. containing spectral information in the $[0-16~kHz]$ band). 
Each branch outputs a decoded time domain signal respectively noted $\hat{d}_{16kHz}$ and $\hat{d}_{32kHz}$. The reconstruction signal of $S_{16kHz}$, noted $\hat{S}_{16kHz}$, is directly obtained from the $16~kHz$ branch, and we have~: $\hat{S}_{16kHz} = \hat{d}_{16kHz}$. The other branch, which operates at $32~kHz$, takes as input the residual signal $S_{32kHz} - U(\hat{d}_{16kHz})$ where $U$ is the upsampling operator\footnote{We use the sinc interpolation function as implemented in \url{https://adefossez.github.io/julius/julius/index.html}.} from $16~kHz$ to $32~kHz$. The reconstruction signal from this branch, noted $\hat{S}_{32kHz}$ is obtained by directly summing the decoded signals from each branch: $\hat{S}_{32kHz} = U(\hat{d}_{16kHz}) + \hat{d}_{32kHz}$.

It is worth highlighting that the proposed cascade architecture enables a form of soft disentanglement. While each branch is primarily designed to generate content within its designated frequency band, no strict constraints enforce this separation. As a result a result, the decoder of the $32~kHz$ branch can also contribute to the reconstruction of lower-frequency content. This soft decomposition allows the $32~kHz$ branch to enhance signal reconstruction below $8~kHz$ and mitigate potential artifacts at the boundaries between frequency bands.

\section{Experiments}
\label{sec-exp}

\subsection{Data}
For our experiments, we used the MUSDB18 \cite{musdb18} and Jamendo \cite{bogdanov2019mtg}  datasets which gather more than 55 000 music tracks in a train set, and 50 music tracks in a test set, initially sampled at $44.1~kHz$ with a total duration above 3 700 hours. Note that MUSDB18 and Jamendo were part of the datasets used in the  training of the original DAC model proposed in \cite{kumar_high-fidelity_2023}.

\subsection{Architecture}
As a base architecture for the disentangled multi-band codec, we chose to replicate the Descript Audio Codec (DAC \cite{kumar_high-fidelity_2023}) architecture, a State of the Art model in audio compression, in a degraded version (i.e. with RVQ containing only few codebooks, or in other words, with a high compression rate). It is composed of a fully-convolutional encoder-decoder architecture, inside of which is plugged a simple Residual Vector Quantizer. 
In our architecture, for each of the two frequency bands we defined earlier, we reproduced a version of this audio codec and adapted its dimensions to the corresponding sampling frequency: we chose the compression rates of the encoding blocks in each branch so that the ratio between the branch sampling rate and the overall compression rate is constant from branch to branch, excluding any latency between the branches.
As for the number of quantizers that are associated with each branch, we decided to reflect the width of each frequency band in the number of tokens that are extracted by the branch. This led to have four quantizers in the $16~kHz$ branch and four other quantizers in the $32~kHz$ branch. 

In order to obtain comparable figures and a fair baseline, we also retrained the DAC model, only on MUSDB18 and Jamendo, at sampling rates $16~kHz$ and $32~kHz$ keeping the compression rates and the bitrate that we defined in our model: a bitrate of $2~kbps$ for the $16~kHz$ model (compression rate of 128) and a bitrate of $4~kbps$ for the $32~kHz$ model (compression rate of 128).
The choice of the DAC structure \cite{kumar_high-fidelity_2023} also makes our model benefit from the same robustness to data diversity.

\begin{figure}[t]
\centerline{\includegraphics[width=1\columnwidth, trim={0 0cm 0 0cm},clip]{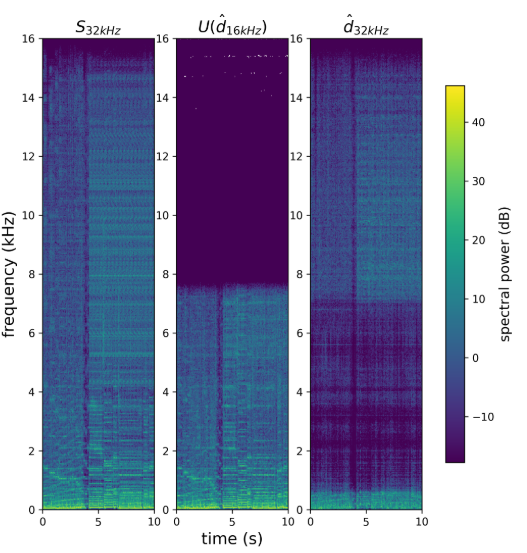}}
\captionsetup{justification=centering}
\caption{Spectrograms of $S_{32kHz}$, $U(\hat{d}_{16kHz})$ and $\hat{d}_{32kHz}$. \\
$U(\hat{d}_{16kHz})$ only encodes information in the $[0-8~kHz]$ band. $\hat{d}_{32kHz}$ has most of its energy in the $[8-16~kHz]$ band, even though it also carries residual information in the lower band.}
\label{fig-spec_dis}
\end{figure}

\subsection{Training Procedure}
As the codec has various branches, a slight adaptation of the training procedure of the DAC model \cite{kumar_high-fidelity_2023} was necessary.
All the branches are first trained in cascade, starting with the branch operating at the lowest sampling frequency: a) we train a branch, b) then freeze its parameters and the ones of all previously trained branches, and finally c) train the next branch. At the end of this cascade training, all the weights of all  branches are finetuned to correctly nest the branches together, and find a global optimum. For each branch, associated to a sampling frequency $F$, we associate a discriminator net  (keeping the structure of the discriminator described in \cite{kumar_high-fidelity_2023}), and we compute adversarial generative ($\mathfrak{L}_{F}^{gen}$) and adversarial feature matching  ($\mathfrak{L}_{F}^{fm}$) losses, a multi scale mel loss ($\mathfrak{L}_{F}^{mel}$) and codebook ($\mathfrak{L}_{F}^{cb}$) and commitment ($\mathfrak{L}_{F}^{cmt}$) losses for each branch.
Let's denote the weighting of these losses $\alpha_{F}^{gen}$, $\alpha_{F}^{fm}$, $\alpha_{F}^{mel}$, $\alpha_{F}^{cb}$ and $\alpha_{F}^{cmt}$.

For the $16~kHz$ and $32~kHz$ first training steps, we define the training losses $\mathfrak{L}_{16kHz}^{total}$ and $\mathfrak{L}_{32kHz}^{total}$:

\begin{equation}
\mathfrak{L}_{16kHz}^{total} = \sum_{\lambda}\left( \alpha_{16kHz}^{\lambda} \mathfrak{L}_{16kHz}^{\lambda}\right)
\label{eq-16_loss}
\end{equation}

\begin{equation}
\mathfrak{L}_{32kHz}^{total} = \sum_{\lambda} \left( \alpha_{32kHz}^{\lambda} \mathfrak{L}_{32kHz}^{\lambda}\right)
\label{eq-32_loss}
\end{equation}
\begin{center}
    where $\lambda \in \{ gen, fm, mel, cb, cmt \}$.
\end{center}
We chose to define the training loss for the finetuning step ($\mathfrak{L}_{finetun}^{total}$), followingly:

\begin{multline}
\!\!\!\!\mathfrak{L}_{finetun}^{total} = \!\!\!\!\!\!\!\!\!\!\sum_{\lambda \in \{ gen, fm, mel \}} \!\! \frac{1}{2} \left( \alpha_{32kHz}^{\lambda} \mathfrak{L}_{32kHz}^{\lambda} + \alpha_{16kHz}^{\lambda} \mathfrak{L}_{16kHz}^{\lambda}\right) \\
+ \sum_{\lambda \in \{ cb, cmt \}} \left( \alpha_{32kHz}^{\lambda} \mathfrak{L}_{32kHz}^{\lambda} + \alpha_{16kHz}^{\lambda} \mathfrak{L}_{16kHz}^{\lambda}\right)
\label{eq-ft_loss}
\end{multline}

Doing so, we compute the mean of the generative, feature matching and mel losses of the two branches, while we simply add the codebook relative losses. These choices are made to account for the difference of nature between the losses which are related to the reconstruction of the target (for which the two branches overlap) and the losses which are constraining the training of the $16~kHz$ and $32~kHz$ codebooks (for which the learned information do not overlap).

Similarly to what we observed in DAC's \cite{kumar_high-fidelity_2023} implementation, we chose for $F \in \{ 16kHz, 32kHz \}$, $\alpha_{F}^{gen} = 1.0$, $\alpha_{F}^{fm} = 2.0$, $\alpha_{F}^{mel} = 15.0$, $\alpha_{F}^{cb} = 1.0$ and $\alpha_{F}^{cmt} = 0.25$.

\subsection{Evaluation}
To evaluate the quality of the reconstruction that is performed by each branch, we also kept the same metrics as in \cite{kumar_high-fidelity_2023}: waveform loss, stft loss, mel loss, scale invariant signal-to-distortion ratio (SI-SDR) as introduced in \cite{le2019sdr}, and the ViSQOL value, a perceptual audio quality assessment introduced in \cite{chinen2020visqol}.

The meaningfulness of the reconstruction was measured thanks to the regular Signal to Distortion Ratio (as the SISDR \cite{le2019sdr} is not fit for band computation) expressed in Decibels, which is defined for $S_F$ a time domain signal sampled at $F$ and its reconstruction $\hat{S}_F$ followingly:
$$SDR(S_{F}, \hat{S}_{F}) =10 \log_{10}\left(\frac{||S_F||^2}{||S_F - \hat{S}_F||^2}\right)$$.

We also conducted an evaluation of the perceptual quality of the reconstructions obtained with our disentangled model, compared to the DAC baseline, through a MUSHRA test \cite{schoeffler_webmushra_2018}. The test required 13 participants to rate from 0 (bad) to 100 (excellent) the perceptual quality of four modalities of a same audio excerpt: a reference, a version encoded by DAC, a version encoded by our model and a degraded version (LP $3.5~kHz$: low pass filter at $3.5~kHz$). This test was performed for both the $16~kHz$ and the $32~kHz$ reconstructions. Two sets of 6 excerpts were drawn randomly from the test set, for both types of reconstructions. Participants were non-experts, primarily researchers in audio processing. All evaluations were carried out using headphones under typical office conditions, with the possibility to replay the audio stimuli as needed.

\section{Results}
\label{sec-Results}

\begin{table}[t]
\caption{Reconstruction metrics of the disentangled codec}
\begin{center}
\begin{tabular}{|c||c|c||c|c|}
\hline
\!\!\!Sampling rate & \multicolumn{2}{|c||}{$16 000~Hz$} & \multicolumn{2}{|c|}{$32 000~Hz$} \\
\hline
\!\!\!Bitrate & \multicolumn{2}{|c||}{$2~kbps$} & \multicolumn{2}{|c|}{$4~kbps$} \\
\hline
\!\!\!Method & \!\!DAC \cite{kumar_high-fidelity_2023}\!\! & \!\!Our model\!\! & \!\!DAC \cite{kumar_high-fidelity_2023}\!\! & \!\!Our model\!\! \\
\hline
\!\!\!mel loss ($\downarrow$)&1.08&\textbf{0.95}&0.90&\textbf{0.80}\\
\!\!\!stft loss ($\downarrow$)&2.67&\textbf{2.52}&2.28&\textbf{2.14}\\
\!\!\!waveform loss ($\downarrow$)&0.072&\textbf{0.066}&0.060&\textbf{0.05}\\
\!\!\!SI-SDR ($\uparrow$)&2.97&\textbf{3.90}&5.00&\textbf{6.05}\\
\!\!\!ViSQOL ($\uparrow$)&4.08&\textbf{4.25}&3.97&\textbf{4.22}\\
\hline
\end{tabular}
\label{tab-reco_met}
\end{center}
\end{table}

\begin{table}[t]
\caption{ \centering Mean perceptual quality scores obtained through the MUSHRA test ($\pm$ std)}
\begin{center}
\begin{tabular}{|c||c||c|c||c|}
\hline
& Reference & DAC \cite{kumar_high-fidelity_2023} & Our model & LP $3.5~kHz$ \\
\hline
$16~kHz$&95 $\pm$ 9&31 $\pm$ 18&\textbf{53} $\pm$ 20&37 $\pm$ 25\\
$32~kHz$&96 $\pm$ 5&49 $\pm$ 20&\textbf{66} $\pm$ 19&38 $\pm$ 24\\
\hline
\end{tabular}
\label{tab-mushra}
\end{center}
\end{table}

\begin{table}[t]
\caption{Disentanglement - SDR by frequency band \\
\textit{[Top] SDR metrics for $\hat{d}_{32kHz}$ per band.}\\ 
\textit{[Bottom] SDR metrics for $\hat{S}_{32kHz}$ around the interface and on the whole spectrogram.}}
\begin{center}
\resizebox{0.3\textwidth}{!}{
\begin{tabular}{|c|c|}
\hline
\multicolumn{2}{|c|}{\rule{0pt}{2.6ex} $\hat{d}_{32kHz}$ on $S_{32kHz}$ \rule[-1.2ex]{0pt}{0pt}} \\
\hline
$[8-16~kHz]$ band & 5.85 \\
$[0-8~kHz]$ band & 3.80\\
\hline
\multicolumn{2}{c}{} \\[2.5ex]
\hline
\multicolumn{2}{|c|}{\rule{0pt}{2.6ex} $\hat{S}_{32kHz}$ on $S_{32kHz}$ \rule[-1.2ex]{0pt}{0pt}} \\
\hline
$[7.9-8.1~kHz]$ (interface) & 5.61\\
Overall mean & 5.67\\
\hline
\end{tabular}
}
\label{tab-SDR}
\end{center}
\end{table}

\vspace{-3pt}

The reconstruction metrics we gathered in Tab. \ref{tab-reco_met} show a slight improvement of the performances of our $16~kHz$ (i.e. using only the $16~kHz$ branch) compared to the baseline. Similarly, the observed reconstruction metrics for the overall $32~kHz$ reconstruction indicate an increase in the quality of the reconstruction compared to the $32~kHz$ version of the DAC model.

The perceptual metrics obtained through the MUSHRA test, and summarized in Tab. \ref{tab-mushra} confirm the observation we made with the reconstruction metrics: for both $16~kHz$ and $32~kHz$ reconstructions, our model performs better than the baseline.

The disentanglement of the information beared by the tokens extracted from our codec was investigated by comparing the output of each branch ($U(\hat{d}_{16kHz})$ and $\hat{d}_{32kHz}$) to the original time domain input signal $S_{32kHz}$. The spectrograms of these signals are displayed in Fig. \ref{fig-spec_dis}.

The reconstruction outputted from the $16~kHz$ branch only encodes information in the low frequencies, as the low sampling rate does not allow for reconstruction above $8~kHz$. As for the reconstruction of the $32~kHz$ branch, most of the information is beared in the high frequencies, even though the soft disentanglement allows for corrections to be made in the low frequency band. We computed a SDR between $S_{32kHz}$ and $\hat{d}_{32kHz}$, splitting the calculation between band $[0-8~kHz]$ and band $[8-16~kHz]$. The results are gathered in the higher part of Tab. \ref{tab-SDR}. 

The signal information reconstructed by the $32~kHz$ branch is, indeed, much more correlated to $S_{32kHz}$ in the higher frequencies (with a SDR of $5.85~db$), even though some relevant information is encoded in the lower frequencies as well (with a lower SDR of $3.80~db$), thanks to the soft disentanglement. 

The results of the lower part of Tab. \ref{tab-SDR} also highlight that the SDR computed in a narrow band around the common cut-off frequency of the two sub-bands ($8~kHz$)  is nearly identical to the average SDR over the full reconstruction band. This shows that the soft disentanglement in our model does not create artifacts at the cut-off frequency.

\section{Application to inpainting}
\label{sec-p_applications}

We demonstrate the interest of our approach by designing a toy experiment addressing the inpainting task, which consists in generating the missing high frequencies from a low-sampled signal.

In that experiment, the disentangled codec takes as input the time domain signals $S_{16kHz}$ and $U(S_{16kHz})$ for the $16~kHz$ and $32~kHz$ branches respectively: it receives no information in the $[8-16~kHz]$ band. As the codec has only been trained with signals containing information in that band, the reconstruction of the $32~kHz$ branch is expected to "fill" the high frequency band, using the information it receives in the $[0-8~kHz]$ band.
We will compare the output of our model $\hat{S}_{32kHz}^{~inp}$, to its reference $S_{32kHz}$. The output of the $32~kHz$ branch is called $\hat{d}_{32kHz}^{~inp}$.

We computed the SDR of each frequency band in $\hat{d}_{32kHz}^{~inp}$, in comparison to $S_{32kHz}$. The results are displayed in Tab. \ref{tab-SDR_inp}.

The SDR value we obtained for the $[0-8~kHz]$ shows the $32~kHz$ branch reconstructs a bit of information in that band.
As for the $[8-16~kHz]$ band result, the SDR indicates that even though the model does not receive high frequencies information, it is still able to interpolate with the data it has in low frequencies to recreate meaningful information (as we see the SDR is slightly above $0~dB$).
This observation is also strengthened by the reconstruction results we show in Tab. \ref{tab-inp_res}.

In this table, we compared $U(\hat{S}_{16kHz})$, $U \circ D(\hat{S}_{32kHz}^{~inp})$ (both of which contain no information in the $[8-16~kHz]$ band) and $\hat{S}_{32kHz}^{~inp}$ to $S_{32kHz}$, $D$ being the downsampling operator from $32~kHz$ to $16~kHz$. 
The better results obtained with the inpainted band, rather than without it, show that the information extrapolated by the $32~kHz$ branch in the setting of that experiment is beneficial to the reconstruction, and bears meaning in relation to ground truth.

\begin{table}[t]
\captionsetup{justification=centering}
\caption{SDR of $\hat{d}_{32kHz}^{~inp}$'s bands compared to $S_{32kHz}$}
\begin{center}
\resizebox{0.28\textwidth}{!}{
\begin{tabular}{|c|c|}
\hline
& \rule{0pt}{2.6ex} $\hat{d}_{32kHz}^{~inp}$ \rule[-1.2ex]{0pt}{0pt}\\
\hline
$[8-16~kHz]$ band&0.54 \\
$[0-8~kHz]$ band&1.16\\
\hline
\end{tabular}
}
\label{tab-SDR_inp}
\end{center}
\end{table}

\begin{table}[t]
\captionsetup{justification=centering}
\caption{Inpainting toy experiment - reconstruction metrics}
\begin{center}
\resizebox{0.5\textwidth}{1.2cm}{
\!\!\!\begin{tabular}{|c||c|c|c||}
\hline
& \rule{0pt}{2.6ex} \!\!\!\!$U(\hat{S}_{16kHz})$\!\! \rule[-1.2ex]{0pt}{0pt} & \!\!$U \circ D(\hat{S}_{32kHz}^{~inp})$\!\!& \!\!$\hat{S}_{32kHz}^{~inp}$\!\!\\
\hline
\!\!mel loss ($\downarrow$)&2.56&2.55&2.04\\
\!\!stft loss ($\downarrow$)&8.65&8.73&5.37\\
\!\!waveform loss ($\downarrow$)&0.067&0.053&0.053\\
\!\!SI-SDR ($\uparrow$)&3.75&6.23&6.22\\
\!\!ViSQOL ($\uparrow$)&2.01&2.01&2.06\\
\hline
\end{tabular}
}
\label{tab-inp_res}
\end{center}
\end{table}

\section{Conclusion}
\label{sec-conclusion}

In this work, we introduce an innovative approach to design a neural audio codec by incorporating a spectral decomposition of input time-domain signals which facilitates feature disentanglement. This method is versatile and independent of the specific data or task at hand. We demonstrate that, with a marginal increase in reconstruction quality, this representation also enhances interpretability in feature extraction. Additionally, our results from a preliminary experiment on inpainting show that this approach offers significant benefits and holds promise for other audio synthesis or sound transformation applications.

\bibliographystyle{IEEEtran}
\bibliography{abrv,bib}

\end{document}